# MODELLING COLLABORATIVE SERVICES:
## *The COSEMO model*

Thanh Thoa Pham Thi[(1)], Thang Le Dinh[(2)]
*(1)School of Computing, Dublin City University, Glasnevin, Dublin 9, Ireland*
*thoa.pham@computing.dcu.ie*
*(2) Université du Québec à Trois-Rivières, Canada*
*Thang.Ledinh@uqtr.ca*

Markus Helfert [(3)], Michel Leonard[(4)]
*(3)School of Computing, Dublin City University, Glasnevin, Dublin 9, Ireland*
*markus.helfert@computing.dcu.ie*
*(4) Department of Information Systems, University of Geneva, Switzerland*
*Michel.Leonard@cui.unige.ch*

Abstract: Despite the dominance of the service sector in the last decades, there is still a need for a strong foundation on service design and innovation. Little attention has paid on service modelling, particularly in the collaboration context. Collaboration is considered as one of solutions for surviving or sustaining the business in the high competitive atmosphere. Collaborative services require various service providers working together according to agreements between them, along with service consumers, in order to co-produce services. In this paper, we address crucial issues in collaborative services such as collaboration levels, sharing data and processes due to business interdependencies between service stakeholders. Afterward, we propose a model for Collaborative Service Modelling – the COSEMO model, which is able to cover identified issues. We also apply our proposed model to modelling an example of healthcare services in order to illustrate the relevance of our modelling approach to the matter in hand.

## 1 INTRODUCTION

Service science is emerged in the last few years as an interdisciplinary research domain that addresses challenges in service innovation in the service sector. The service sector includes all economic activities whose output is not a tangible product and is generally consumed at the same time, it is produced and provides added value in intangible forms (Quinn et al, 1987). Today, more and more business organizations have been seeking collaborations as one amongst solutions to sustain their business in high competitive environments. For instance, they use the supply chain model or alliance with partners. In this context, *collaborative service* is a kind of business collaboration in the service sector in which several business organizations work together to co-produce services.

Besides, although the dominance of service sector in recent years, the strong foundation on service design and innovation is still needed (Bitner et al, 2008). Service modelling is undertaken in the service design and innovation. Service modelling amounts to the representation of relations between what is provided to customers, how it is provided, the technical definition of the service, and resources needed for operating the service (Vilho Raisanen, 2006). Therefore at the informational level, service modelling should describe the creation and transformation of information between service stakeholders.

In the collaboration context, a service model should show the business aspect of the collaboration such as the participant role, sharing and exchanging of information, common processes, and the collaboration degrees.

Actually, approaches for service modelling often focus on modelling a single service (Shostack, 1982), (Kingman-Brundage, 1995), (Vilho Räisänen, 2006), (Zeithaml, 2008). There are few approaches working on inter-organizational business process collaboration such as (BPMI, 2004), (Grossman, 2008). In our point of view, they need some adaptations and extensions for modelling collaborative services.

In this paper, we point out the principles of collaborative services and then we present a new model for collaborative service modelling. We observe that, at the informational level, the collaboration degree between various business organizations is based on interdependencies of data and processes between these organisations, which are represented by the needs of sharing of data and processes. Consequently, the data consistency and correct performing of shared processes in the collaboration context should be taken into account. Our approach therefore exploits these issues.

The rest of the paper is organized as follows: Section 2 dues with concepts of service and collaborative services; it also discusses some related works Section 3 clarifies characteristics of collaborative services such as collaboration level and interdependency issues across organizations. Section 4 presents our proposed model for collaborative service modelling. Section 5 describes an application of our modelling approach to an example of healthcare service. Finally, we conclude our work and give some research directions in Section 6.

## 2 BASIC CONCEPTS AND RELATED WORKS

### 2.1 Services and collaborative services

Services are defined as "the application of competences for the benefit of another, meaning that service is a kind of action, performance or promise that is exchanged for value between provider and client" (Spohrer, 2007).

Basically, there are four main characteristics of services (Tien, 2003):
- Information-driven: the creation, management and sharing of information is essential to the design and the production of services;
- Customer-centric: customers are co-producers of services that can require the adaptation or the customization of services;
- Electronic-oriented: the achievements of information and communication technologies improve the automation and the connection between economic entities and enable e-commerce, e-business, e-collaboration, e-government and then e-services; and
- Productivity-focused: in order to obtain the competitive advantage in the global economy, services are measured based on dimensions of performance measurement such as efficiency and effectiveness.

These characteristics of services require the service description should focus on creation, sharing and exchanging of information (i.e. interaction) between service providers and service customers. In other words, "management of services is closely related to processes, or the order in which different tasks required for creating and operating a service are carried out" (Vilho Raisanen, 2006).

Collaborative services in our context are services demanding collaboration of service co-providers which are various economic entities, together with service client to co-produce services.

According to (IEC, 2005; Kosanke, 2005; Touzi et al, 2009), there are different levels of collaborative maturity that an organization can adapt: *communicating* (capable of exchanging and sharing information), *open* (capable of sharing business services and functionalities with others), *federated* (capable of working with others according to a set of collaborative processes that have a common objectives and to ensure its own objectives) and *interoperable* (capable of working together without a special effort, partners appear as a homogeneous and seamless system). For instance, there is a need of collaboration between transport service providers, accommodation service providers, sightseeing tour provider to produce travel package services; these organizations may adapt *federated* collaborative level. Meanwhile the collaboration between hospitals, general practitioners, and medical test laboratories in the production of healthcare services may adapt the *communicating* collaborative level.

In our point of view, in the context of collaboration, the sharing and exchanging of business information and business processes between organizations are essential issues irrespective of collaboration level. Furthermore, the consistency of business information and processes between business partners must be ensured. Thus the modelling of collaborative services should focus on describing various degrees of sharing and exchanging of business information and processes and their consistency according to the collaboration levels between business partners.

## 2.2 Services and collaborative services modelling

Most of current approaches for service modelling focus on the interactions and transformation of information between service provider and service consumers. Meanwhile modelling the collaboration between service providers received little attention, in terms of description of data sharing and business processes sharing.

*Molecular model* is one of the earliest models for service and product modelling developed by Shostack (Shostack, 1982). This model considers "the total market entity" as an atom. The centre of molecular model describes core benefit provided to customers. The core benefit includes *service elements, product elements, relationships* between elements, and *service evidences*. These elements can be visualised with graphical notations.

*Service elements* are services purchased or used by customer (e.g. caring service). A service element often links to physical objects which are evidences of service existence or completion of service, so-called *service evidences* (e.g. medicine), the linkage is the *relationship* between service elements and service evidences. Other elements such as *price strategy*, *distribution strategy*, and *advertisement strategy*, etc. are layers outside the core of the atom describing the total market entity. However, the *molecular* model does not illustrate how the service functions, but it describes the guideline for offering a service.

The *blueprinting* is another approach for service modelling also developed by Shostack (Shostack, 1984) and then evolved by (Kingman-Brundage, 1995), (Vilho Räisänen, 2006), (Zeithaml, 2008) which has focused on processes that constitute the service. A service blueprint is a two-dimensional diagram. The horizontal axis represents the chronology of *processes* or functions in the service. The process handling such as failure point, bottleneck is also described. The vertical axis represents different areas of processes such as visible part for describing *service evidence*, the *front stage* area for the presentation of processes having interactions with customer (visible to customers) and the *back stage* area for the presentation of supporting processes (invisible to customer). The blueprint approach aims at modelling services, but not collaborative services; therefore elements concerning collaborative process have not been mentioned yet.

Recently, the Business Process Modelling Notation (BPMI, 2004) has been adapted for collaborative business process modelling for the Service Oriented Architecture design (Touzi et al, 2009). The collaboration process or global process describes interactions between two or more business organizations. A*ctivities* concept represents tasks or sub-processes within the collaboration process. *Activities flow* may be categorized as *sequence*, *conditional*, *exception*, etc. within *gateway type* as OR, XOR, Parallel, Fork, Join. *Data* represents input/output information of activities. The *message flow* concept represents data exchange or data flow between *actors* or participant business organizations. Message flow and activities flow may be handled with *start event*, *intermediate event*, or *end event*.

A *collaborative process* is described in a *pool*, which may be divided to several *lanes*. Each *lane* depicts interactions of each participant in the collaborative process.

Also, (Grossman, 2008) has proposed some extensions to business process modelling languages which enable them to describe various types of inter-process dependencies across organizations. They are *Triggering dependencies*, *Enabling dependencies*, *Cancelling dependencies*, *Disabling dependencies*.

These approaches can be adapted to model exchanging and sharing of information between service stakeholders. But they need some extensions to describe systematically the collaboration degrees, monitoring data consistency and monitoring shared process performing across organizations.

## 3 ISSUES ON MODELLING COLLABORATIVE SERVICES

This section addresses the important challenges in collaborative services modelling.

## 3.1 Collaboration level

We review the collaboration level of service co-providers based on data and process interdependencies perspective rather than the business strategy perspective as mentioned in Section 2.1 (IEC, 2002; Kosanke, 2005, Touzi et al, 2009). By this way, the collaboration level is applied to each common/sharing process and data. Thus between two organizations, there may be a mix collaboration level depending on every sharing process and/or data. This flexibility is well practical, because important degrees or pivotal degrees of business processes and data are various in an organization, therefore their interdependencies levels are also different.

Data and process interdependencies in the inter-organizational collaboration are represented by data and process sharing needs. So, we propose four collaboration levels as follows.

- Level 1 – *very tight* collaboration: a service provider shares processes and information to other service provider. It is allowed to modify the common information. There is no explicitly sharing information request, but it is made automatically for the correct function of shared processes.

  For instance, a hotel collaborates with a travel agency who provides booking service to customer. Customers can contact the travel agency for their booking or cancellation, or they can also book or cancel their reservation directly with the hotel. Of course, a booking is made with the hotel just be changed or cancelled with the hotel, a booking is made with the travel agency may be changed or cancelled with the travel agency or with the hotel. The shared information between the hotel and the travel agency are Room situation, Booking, and Cancellation. Customer information owned by travel agency should be shared to the hotel, but the hotel does not need to share information of Customer who directly booked with the hotel, to the travel agency. In this case, the travel agency and the hotel share the booking process and some information, and the hotel can modify information shared by the travel agency (e.g. modify check-in date or cancel a booking)

- Level 2 – *tight collaboration*: service providers share processes and information to each other for the reference purpose. It is not allowed to modify the common information.

  For instance, an alliance of airlines shares the process of *Bonus Calculation* for loyalty clients to each member. The shared information between them for the performing of this process is information on flights taken, miles and customer. However, an airline can not modify this shared information which is owned (created) by other airlines.

- Level 3 – *loose* collaboration: a service provider shares information to other service provider according to the request of the first one, and the waiting time is critical to the requesting service provider. This case often concerns requests for carrying out processes, and the shared information is the result of that process realisation.

  For instance, a general practitioner (GP) sends a patient to a laboratory for performing a scanning test; the GP shares patient information, and scanning request to the laboratory; in turn the laboratory shares to GP the scanning result, the GP can not proceed his process without the scanning result.

- Level 4 – *very loose* collaboration: a service provider shares some information to other service provider for the reference purpose, the waiting time of received information is not important.

  For instance, a GP sends a patient to a hospital for further treatment. The GP shares to the hospital information about the patient and his diagnosis. This information is considered as input information for further treatment in the hospital. Once the patient finished his treatment at the hospital, the hospital sends a report to the GP concerning patient situation for further reference. In this case, both hospital and GP did not explicitly request information, thus for them the waiting time does not affect their processes.

Table 1 summarizes the four proposed collaboration levels.

Table 1: Collaboration level of service co-providers

| Collaboration level | Information and process sharing |
|---|---|
| Very tight | Sharing information and process, and allowing to modify sharing information |
| Tight | Sharing information and process, but not allowing to modify sharing information |
| Loose | Sharing information for reference, waiting time is important |
| Very loose | Sharing information for reference, waiting time is not important |

## 3.2 Consistency of data and process

Consistency issues emerged in the context of collaborative services due to data and process interdependencies among service providers.

In general, a process receives input information and produces output information; in some cases a process transforms information from a state (as input information) to another state (as output information). In the latter case, there are two situations: there is an exclusion of these two states in the business context (e.g. *occupied room* and v*acant room*) or there is not (e.g. *occupied room* and *cleaned room*). If there is no exclusion, then information remains in both states, otherwise information is changed completely. Let's illustrate this issue with the below example (Figure 1). A small private hospital collaborates with a cleaning agency. The hospital shares information on *occupied room* and *vacant room* to the cleaning agency. The agency offers two kinds of cleaning room service: normal cleaning for occupied room and total cleaning for rooms just have been vacant.

Figure 1: Example on process constraint between service providers

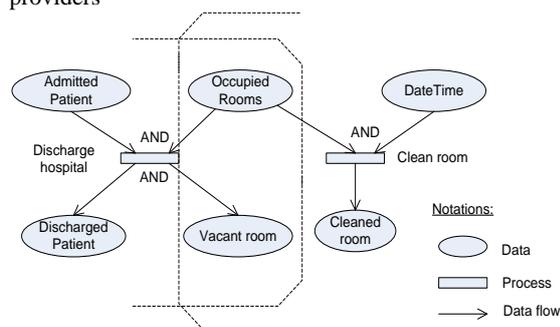

An *occupied room* is transformed completely to a *vacant room* when the patient staying in this room is discharged from the hospital, but an *occupied room* remains always occupied after it is cleaned (normal cleaning) as long as the patient has not been discharged from the hospital yet. In other words, there is a constraint on processes between partners in this case. How can we describe this constraint in collaborative service modelling: "For every room, after the *discharge hospital* process is performed then the *clean room* process is disable, but it is not the case on the contrary"? Our approach allows to describe this with the concept of remaining state and leaving state of information object.

Let's suppose there is a token in the *Occupied room* class, after the token participates in the *clean room* process, a new token is created in the *Cleaned room* class and the token in the *Occupied room* class is still remained. Meanwhile if the token in *Occupied room* class participates in the *discharge hospital* process, it will leave the *Occupied room* class and come to the *Vacant room* class, thus this token cannot any more participate in the *clean room* process.

We take again the example on collaboration between a hotel and a travel agency for illustrating another situation. Apart from *Made booking* shared process, the hotel has its own *Check-in* process, and the travel agency has its own *Cancel* process. The shared data between them are *Customer*, *Booking*, *Cancelled booking* (Figure 2).

In this example, if a booking is cancelled at the travel agency, then it must be communicated or shared to the hotel with the *cancelled booking* information, so that the concerning customer cannot check-in the hotel with this booking. At the travel agency side, if there is a *confirmed cancel request*, then at the hotel side there can not be a *checked-in customer* concerning this booking. In the same way, this data constraint between partners can be described with the remaining state/leaving state concepts: a *booking* object completely transformed to *cancelled booking* state (leaving the *booking* state) when it participates in the *Cancel* process, thus it can not any more participate in the *Check-in* process, and similarly in the case *check-in* a booking before cancelling it. Therefore it also represents a process constraint. This case clearly illustrates the case where process constraint implies data constraint.

Figure 2: Example on data constraint and process constraint between service providers

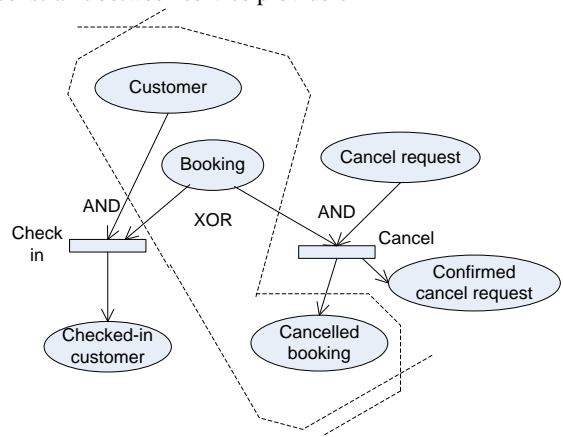

In this section we have discussed critical issues about collaboration levels and interdependencies between business organizations in the context of collaborative services. The next section presents our proposed model for collaborative service modelling taking into account these issues.

## 4 THE COSEMO MODEL

The **Co**llaborative **Se**rvice Conceptual **Mo**delling (COSEMO) model is inspired by the IASDO model (Pham-Thi et al, 2006, 2008). The COSEMO model is capable of firstly covering issues discussed in the previous section, and secondly ensuring the consistency of the modelling concepts for obtaining a consistent model.

The COSEMO model includes the following concepts.

- C: represents a set of classes c of information objects created, transformed during the service realization. An instance of C may also be a significant *dynamic state* of an information object, which is transformed from another state through a process realisation;

- P: set of business activities or process involved in the service. A process creates output information and/or transformed the input information to its new state as output information;

- R: set of organizational roles or partners who are responsible for process execution in service. Partners may also own business processes in their organization. Normally a process is owned by an organization. In the collaboration context, it may be owned by several organizations;

- fi: (C, P) → {0,1}, function of input information, if fi(c, p)= 1, c ∈ C, p ∈ P, then c is an input class of p;

- fo: (P, C) → {0,1}, function of output information, if fo(p,c)= 1, c ∈ C, p ∈ P, then c is an output class of p,

- fsp: ({*Waiting point, Fail point, Decision point*}, C) → {0,1}, function of status points, status points are particular characteristics of input/output information which is significant in service realization, the status points are attached to classes, , e.g. if fsp('Waiting point', c)=1, c ∈ C, then c is a waiting point;

- f_remain: (C, P, C) → {0,1}, if f_remain (c1, p, c2)=1, fi(c1, p) =1, fo(p, c2)= 1 then the c1 state is remained when c1 transformed to c2 state through p realization, otherwise if f_remain (c1, p, c2)=0, fi(c1, p) =1, fo(p, c2)= 1 then the c1 state completely transforms to the c2 state;

-p_privilege: (R, P) → {*owner, responsibility*}, defining roles as owner of a process or in charge of a process e.g. if (r, p)= *owner*, then r is an owner of the p process;

-c_privilege: (R, C) → {*creation, modification, reference, suppression, modification$^+$, reference$^+$, suppression$^+$*}\*, assigning privileges on data to roles. *Modification, reference,* and *suppression* privileges are applied on the role's own data. *modification+, reference+,* and *suppression+* are privileges applied on data created by other roles, e.g. if c_privilege(r,c)={*creation, modification, modification+*} then the r role can create instances of c class, and r can modify instances in the c class created by c and other roles.

The COSEMO model is capable of describing collaboration levels between two service providers as follows.

Level 1 – *very tight collaboration*: Suppose r1 and r2 are service co-providers, p is sharing process, c is sharing information which is an output information of p, then fo(p, c)= 1, p_privilege(r1, *owner*)=1, p_privilege(r2, *responsibility*)=1, c_privilege(r1, c)={ *creation, modification, reference, suppression, modification+, reference+, suppression+*}, c_privilege(r2, c)={ *creation, modification, reference, suppression, reference +*}.

Level 2 - *tight collaboration*: fo(p, c)= 1, p_privilege(r1, *owner*)=1, p_privilege(r2, *owner*)=1, c_privilege(r1, c)={ *creation, modification, reference, suppression, reference+*}, c_privilege(r2, c)={ *creation, modification, reference, reference +*}.

Level 3 - *loose collaboration*: suppose r1 service provider shares c information to r2 service provider and the waiting time of r2 for receiving c is important, then c_privilege(r1, c) ={*creation, reference, ...*}, c_privilege(r2, c)={*reference+*}, and fsp(c, *waiting point*)=1.

Level 4 - *very loose collaboration*: it is the same as level 3 modelling, but the waiting time of r2 for receiving c is not important, then c_privilege(r1, c) ={*creation, reference, ...*}, c_privilege(r2, c)={*reference+*}, and fsp(c, *Waiting point*)=0.

As mentioned above, our model is also capable of describing the process and data constraints due to interdependency issues thanks to concepts of remaining state or leaving state of information object (f_remain function).

Modelling with the COSEMO model may be visualized with a graphical data-process-organizations diagram, in which an oval shape represents a class, a rectangular represents a process, swim-lanes represents organizational units or partners. The modelling process can be completed with definitions of functions c_privilege, p_privilege, f_remain.

Figure 3 depicts the meta-model of the COSEMO model with the UML class diagram notations (Rumbaugh et al, 1999).

The *Dynamic State* class describes information in a specific state, which is significant to the business process. It is up to the modeller to decide dynamic states. Output *data flow* class describes an output class of a process, the *Input data flow* class describes an input class of process. Besides, inspired by the *blueprint* approach, we identify status points during the service production which are *Fail point*,

*Decision point* and *Waiting point*. These points are attached to information or state. A *Fail point* describes where the service meets obstacle; there should be a backup solution to resolve this problem. A *Decision point* presents a need of decision or a choice in the service production. A *Waiting point* links to shared information that an organisational unit or a partner is waiting for.

There are some constraints in the meta model, in order to ensure a consistent modelling. These constraints are as follows:

(C1) classes in *Input data flow* class and *Output data flow* class which are involved in the *Remaining state* class must belong to *Dynamic state* class.

(C2) If a role r has the *creation* privilege on a class c then it also has the *reference* privilege on c.

(C3) If a role r has *owner* or *responsibility* privileges on a process p, then r has *reference* privilege on all input classes of p.

(C4) If a role r has *owner* or *responsibility* privileges on a process p, then r has *creation* privilege on all output classes of p.

(C5) If a role r has *owner* privilege on a process p, then r has *reference+* privilege on all input class and output classes of p.

Figure 3: Meta-model of the COSEMO model

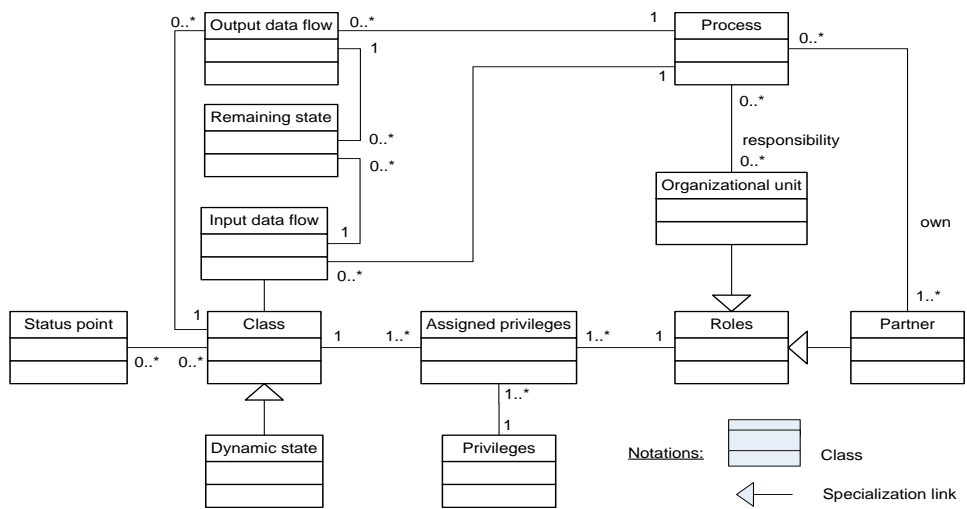

## 5 EXAMPLE

We illustrate our modelling concepts with the example of collaborative healthcare services. Economic entities concerning the collaborative healthcare service in this example are patient, general practitioners (GP), hospitals, and private laboratories that perform medical tests such as blood test, x-ray scan, ultra-sound, etc. They work together to provide the best healthcare services to patients. Apart from local agreements set up between them, their collaboration is also regulated by the country laws.

A patient just admits the hospital in urgent cases; otherwise he/she must consult the GP first. Basing on the patient health situation, the GP decides to treat the patient or to send him/her to the hospital. In the first case, the GP may also request the patient to do some laboratory tests before caring. In the second one, the GP must give a letter of recommendation to the patient for treating at a hospital. The patient then admits the hospital with the GP letter. When the patient is discharged from the hospital, the hospital sends the GP a report about the patient caring situation for reference.

The modelling of this collaborative service is presented in Figure 5.

In this model, decision points are *Diagnosed Patient* and *Check-up patient* which mean that at this state, there is a need of making decision about what next process should be undertaken. The decision corresponds to an alternative process execution. For instance, after diagnosing the patient (i.e. at the state of Diagnosed Patient ) the GP must decide either recommend the patient to the hospital, or request some laboratory tests, or take care the patient right after.

Waiting point in this example is *Sent test result*.

Fail points in this modelling include *Report of patient* and *Sent test result*. The laboratory must send the test result to the GP in order to suitably caring the patient. In case the GP does not receive the result because of a delivery mistake, then the service is stalling.

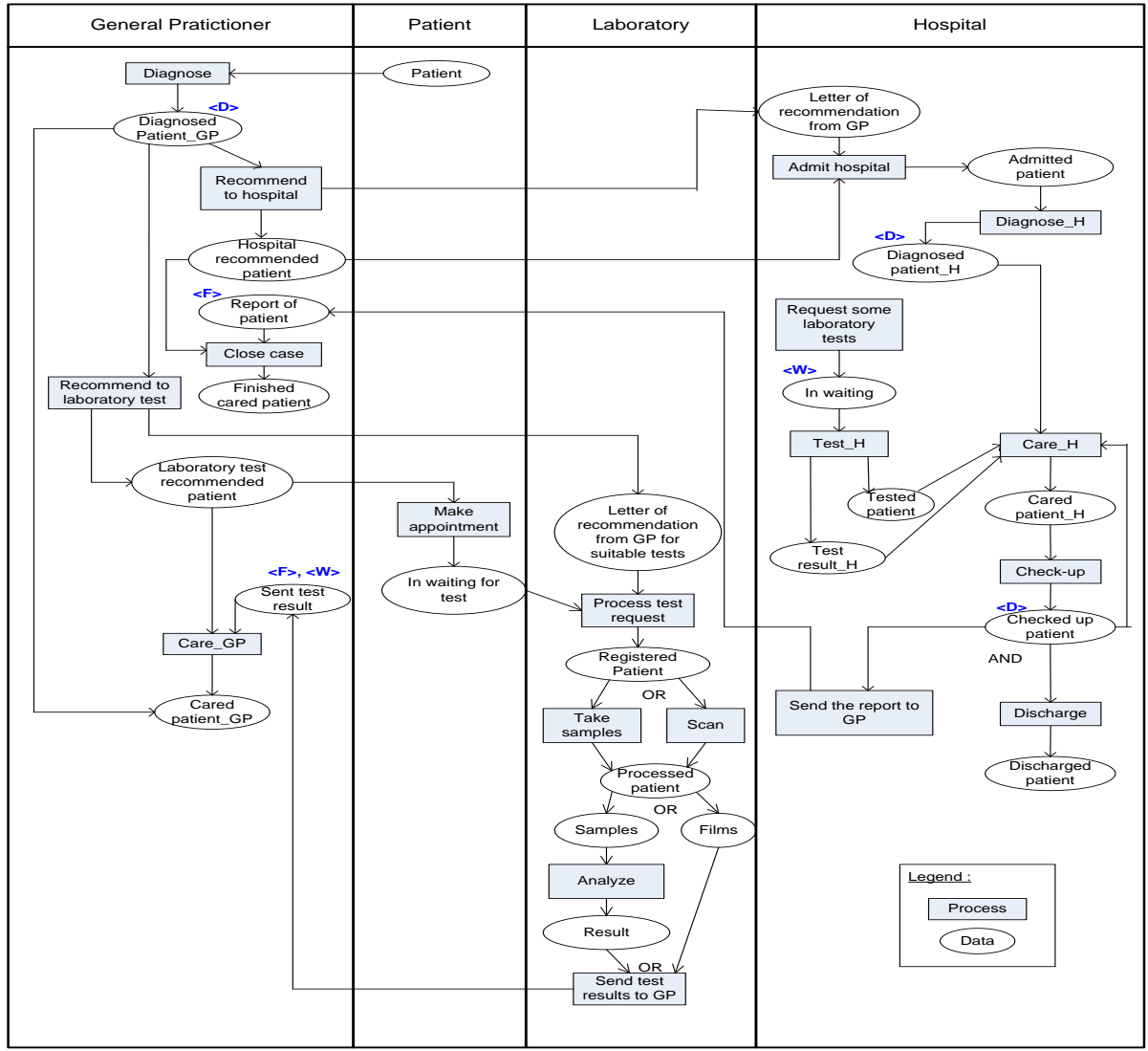

Figure 4: Healthcare service modelling

In this case, the service designer should clarify a sub-process or some agreement between Laboratory and GP to overcome this situation, for instance the result letter should always be a registered letter.
The information sharing between service co-providers is clearly described in this example, for instance the GP shares Letter of recommendation to Laboratories, in turn Laboratory shares test result to GP. Basing on the sharing information, it is said that the collaboration level between GP and Laboratories is a loose collaboration, the collaboration level between GP and hospital is a very loose collaboration. There is no constraint on data and processes between service providers in this example.

Because of the limit of space, in this paper, we don't model this example with other current approaches and compare to our modelling. However, as discussed in Section 2, our approach is able to describe issues that are not covered in other approaches. Our model is suitable for modelling at the informational level. Therefore, it is independent of technical issues, such as Electronic Data Interchange technology implementation for data sharing purpose.

# 6 CONCLUSIONS

Since the last few years, service sciences have attracted attention of many researchers. However, there is still a need for a strong foundation for service design and innovation. Particularly, service modelling and collaborative service modelling have received little attention from researchers. Collaborative services are services which demand collaboration between various service providers to co-produce services. This paper contributes to that research gap with the following:

Firstly, it proposes an approach for collaborative service modelling in which the collaboration concepts are clarified basing on the exchanging and sharing information and process between collaboration partners or service providers. Four collaboration levels are identified according to the sharing degree of information and processes and the privileges of partners on shared information and processes. Besides, the sharing of data and processes represents interdependencies of service providers. As a consequence, constraints on data and processes are emerged in order to ensure consistency of data and processes between service providers.

The paper then presents the COSEMO model for collaborative service modelling. Concepts of dynamic state, status point, data privilege, process privilege and remaining state of information object allowed to cover identified collaborative issues. Furthermore, the model ensures a consistent modelling between modelling concepts.

The case study at the end of this paper also shows that our approach is pertinent in term of providing a relevant modelling approach for collaborative service at the informational level.

Concerning our future research directions, there is a need of investigation on the customer-centric characteristic of services which allow the service customization and adaptation of customers according to their needs. Therefore the modelling should be an adaptable approach and facilitate evolution. Concretely, we analyze adaptation situations of services and then the expression-ability of the COSEMO model will be revised and adapted as well.